%% file: srw.tex
\documentclass[aps,prstab,twocolumn,amsmath,amssymb]{revtex4}

\usepackage{graphicx} 
\usepackage{dcolumn}  
\usepackage{bm}       

\usepackage{makeidx}
\makeindex

\newcounter{one}
\newcounter{two}
\begin{document}


\title{
  Estimation of
  synchrotron-radiation background 
  based on a real beam orbit \\
}


\author{Tetsuo Abe}
\email[E-mail address: ]{tetsuo.abe@kek.jp}
\affiliation{Accelerator Laboratory,
             High Energy Accelerator Research Organization (KEK),
             Tsukuba, Ibaraki 305-0801, Japan 
            }
\author{Hitoshi Yamamoto}
\affiliation{Research Center for Neutrino Science (RCNS), Tohoku University,
             Sendai, Miyagi 980-8578, Japan 
            } 

\author{ }
\affiliation{ }

\date{\today}

\begin{abstract}
Some high-energy experiments have suffered from 
  synchrotron-radiation background.
As a measure,
  we have developed a new calculation method of synchrotron radiation
  based on a real beam orbit,
  aiming at quantitative estimations 
         and the construction of a possible alarm system for the background.
An explanation and a demonstration of our method are given.
\end{abstract}


\maketitle


\input{Introduction/introduction}

\input{Orbit_calc/orbit}

\input{Wattage_calc/wattage}

\input{Benchmark/btest}

\input{conclusion}
\appendix
\input{Appendix/transfer_matrices}

\begin{acknowledgments}
We are grateful to
  Y.~Funakoshi, S.~Kamada, K.~Kanazawa, M.~Kikuchi, H.~Koiso, M.~Masuzawa,
    T.~Nakamura, K.~Oide, R.~Sugahara, N.~Yamamoto,                   
    J.~Haba, S.~Stanic, S.~Swain, K.~Trabelsi, T.~Tsuboyama, S.~Uno,  
    M.~Yokoyama, Y.~Namito,
    and Y.~Yamada 
    for the fruitful comments and discussions.
\end{acknowledgments}

\bibliography{srw}


\end{document}

%% file: Introduction/introduction.tex
\section{\label{R:Sec:intro}
  Introduction}
At the initial stage of the KEK B-factory (KEKB) experiment \cite{C:KEKB:DesignReport95},
  synchrotron radiation (SR)
    from the circulating $8$-GeV electron beam
  caused serious damage to the Belle silicon vertex detector (SVD) \cite{C:Belle:SVD1},
   which was located close to the interaction point (IP).
Some of the preamplifiers mounted on the inner-most SVD layer
  died in 
  about ten days just after start of the gain drop.
%
Other high-energy experiments have also suffered from SR background.


We have developed a new method
  to calculate SR wattages 
  based on a real beam orbit,
  aiming at quantitative estimations 
            and the construction of a possible alarm system for SR background.
Here, the real orbit is obtained
  by fitting measurements of beam-position monitors (BPMs).
In the following sections,
  our method
  is explained 
  together with some illustrations using the $8$-GeV electron beams
  of KEKB High Energy Ring (HER).
%
Finally, we present a simulation of the orbit and the gain drop
  at the time of the SVD gain-drop accident
  as a bench mark.


%% file: Orbit_calc/orbit.tex
\section{\label{R:Sec:orbit_calc}
  Orbit calculation}
%
Beam orbits are calculated by a linear approximation
  using transfer matrices, 
  which 
  are shown in Appendix \ref{R:App:TransferMatrices}.
For numerical calculations, we coded a dedicated computer program in Fortran,
  where
  the correctness of the orbit calculation 
  was checked by comparing with a calculation using SAD \cite{C:KEKB:SAD}.
Furthermore, we confirmed
  that there was no significant effect from round-off errors 
  by comparing the results
    of double- and quadruple-precision calculations.

\subsection{Fitting procedure}
Orbit fitting is performed
  based on the optics, field strengths of the magnets, and BPM measurements.
We require minimization of $\chi^2$, defined as
{\small
\begin{eqnarray}
  \chi^2  &=&
        \sum_{j:{\rm BPM}} \bigg\{
          \Bigl( X_{j}^{\rm (BPM)} - X_{j}^{(orbit)} \Bigr)^2 \nonumber\\
       & & + 
          \Bigl( Y_{j}^{\rm (BPM)} - Y_{j}^{(orbit)} \Bigr)^2
        \bigg\} ~/\sigma_j^2 ~{\rm ,}
  \label{E:OrbitCalc:chi2}
\end{eqnarray}
where
  $X_{j}^{\rm (BPM)}$ ($Y_{j}^{\rm (BPM)}$) indicates
    the horizontal (vertical) beam position measured with BPM:$j$,
  $X_{j}^{\rm (orbit)}$ ($Y_{j}^{\rm (orbit)}$) 
    the horizontal (vertical) position of the calculated orbit at BPM:$j$, and
  $\sigma_j$ is the BPM:$j$ resolution.
There are four floating parameters, which are
  the orbit positions at the entrance and exit
  of the relevant accelerator section:
  $X^{(in)},Y^{(in)},X^{(out)},Y^{(out)}$.
Minimization of $\chi^2$ is performed using the computer program MINUIT
   \cite{C:MINUIT:CERNLIB,C:MINUIT:1975}.
An example of the orbit-fitting results for KEKB HER
  is shown in Fig.~\ref{F:OrbitCalc:fit_no_corr}.
Here, we define the vertical and horizontal $\chi^2$ 
  to estimate the goodness of the fit for each direction separately:
\begin{eqnarray}
  \chi_X^2  &=&
        \sum_{j:{\rm BPM}}
          \Bigl( X_{j}^{\rm (BPM)} - X_{j}^{(orbit)} \Bigr)^2 ~/\sigma_j^2
          ~{\rm ,} \label{E:OrbitCalc:chi2_X} \\
  \chi_Y^2  &=&
        \sum_{j:{\rm BPM}}
          \Bigl( Y_{j}^{\rm (BPM)} - Y_{j}^{(orbit)} \Bigr)^2 ~/\sigma_j^2
          ~{\rm ,} \label{E:OrbitCalc:chi2_Y}
\end{eqnarray}
  which lead to the following relations:
\begin{eqnarray}
        \chi^2  &=& \chi_X^2 + \chi_Y^2 ~{\rm ,} \\
        \frac{\chi^2}{N_{df}} 
                     &=& \frac{{\frac{\chi^2_X}{N_{df(X)}}}\frac{1}{N_{df(Y)}}
                          +{\frac{\chi^2_Y}{N_{df(Y)}}}\frac{1}{N_{df(X)}}}
                                {\frac{1}{N_{df(X)}}+\frac{1}{N_{df(Y)}}}
  \label{E:OrbitCalc:chi2_X_Y} ~{\rm ,}
\end{eqnarray}
  where
    $N_{df(X)}$ ($N_{df(Y)}$) indicates the number of degrees of freedom related to
    the horizontal (vertical) direction, and
    $N_{df}$ for both: $N_{df} = N_{df(X)}+N_{df(Y)}$.
In the example shown in Fig.~\ref{F:OrbitCalc:fit_no_corr},
  the goodness of the fit is
  $\chi_X^2/N_{df(X)}=10.5$ and $\chi_Y^2/N_{df(Y)}=40.3$
  for a conservative BPM resolution of $100$\,$\mu$m. 
%
As is often the case, $\chi^2$ is too large, and
  also the orbit at the IP is unrealistically deflected
    in the negative direction by about $3$\,mm.
This is due to the fact that the absolute positions of magnets and BPMs are
not known sufficiently well, leading to a need for the offset corrections.

\begin{figure}
  \centering
  \includegraphics[width=42mm,clip]{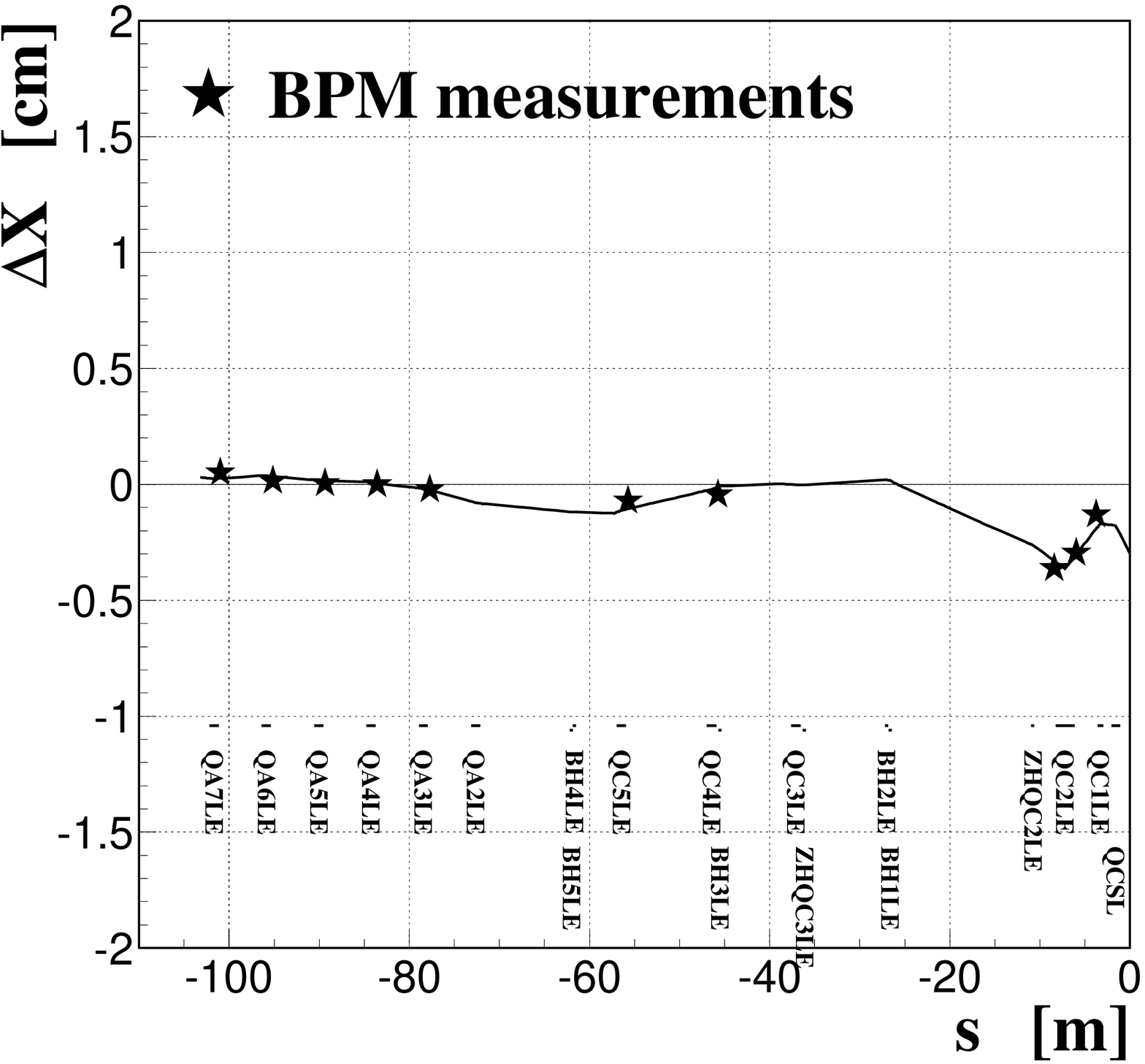}
  \includegraphics[width=42mm,clip]{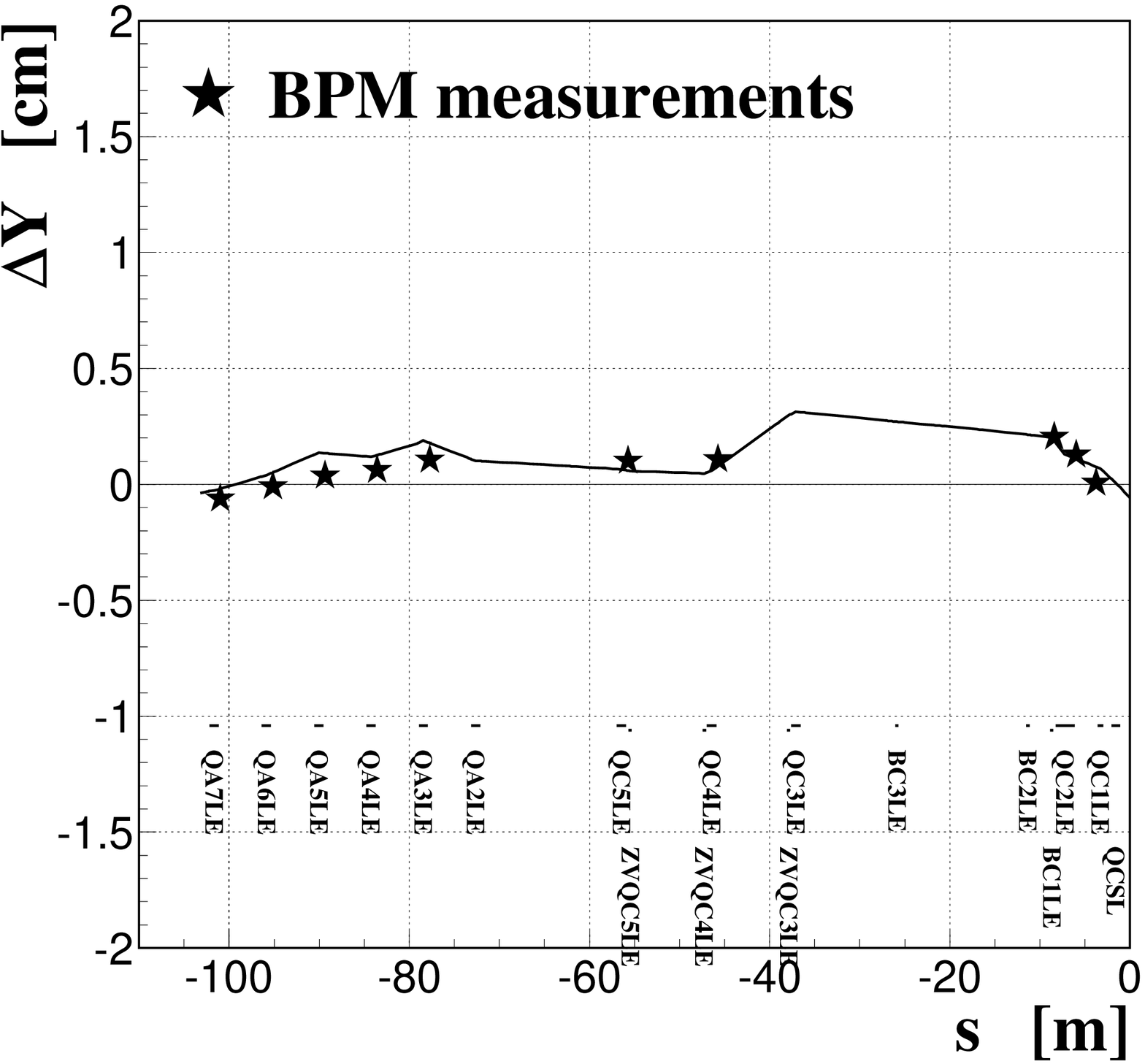}
  \caption{Example of the orbit-fitting results
           for KEKB HER around the IP ($s = \Delta X = \Delta Y = 0$)
           without any correction.
           The left (right) figure shows the horizontal (vertical) fitted orbit
           together with the BPM measurements.
           The $s$ axis is taken to be along the orbit.
           The horizontal line with $\Delta X = \Delta Y = 0$
             corresponds to the design orbit.
           The HER beam flows from the left to the right.
           $\chi_X^2/N_{df(X)}=10.5$ and $\chi_Y^2/N_{df(Y)}=40.3$.
          } \label{F:OrbitCalc:fit_no_corr}
\end{figure}

\subsection{Offset corrections}
%
We modify the $\chi^2$ formula~(\ref{E:OrbitCalc:chi2}) as follows:
{\small
\begin{eqnarray}
\chi^2 &=& \sum_{i:{\rm{orbit}}}^{ }
     \sum_{j:{\rm BPM}}^{ }
           \biggl[               \nonumber\\
     & &
     \Bigl\{ 
           \bigl( X_{j,i}^{\rm (BPM)}
                       + {\Delta X_{j}^{\rm (QUAD)}}
                       + {\Delta X_{j}^{\rm (BPM)}}
                   \bigr) - X_{j,i}^{(orbit)} \Bigr\}^2 \nonumber\\
    &+&
     \Bigl\{ 
             \bigl( Y_{j,i}^{\rm (BPM)}
                     \,+ {\Delta Y_{j}^{\rm (QUAD)}}
                     \,+ {\Delta Y_{j}^{\rm (BPM)}}
                   \bigr) - Y_{j,i}^{(orbit)} \Bigr\}^2 \nonumber\\
           & & \biggr] ~/\sigma_j^2 ~{\rm ,}
  \label{E:OrbitCalc:chi2:offsets}
\end{eqnarray}
}
  where
    $\Delta X_{j}^{\rm (QUAD)}$ ($\Delta Y_{j}^{\rm (QUAD)}$) indicates
    the horizontal (vertical) offset, with respect to the design position,
    of the real field center of the quadrupole magnet (QUAD)
      on which BPM:$j$ is mounted,
  and $\Delta X_{j}^{\rm (BPM)}$ ($\Delta Y_{j}^{\rm (BPM)}$) 
    the horizontal (vertical) offset of the real origin
      of the BPM:$j$ measurement
    with respect to the real field center of QUAD:$j$.
The index $i$ runs for different orbits in the case of a global fitting,
  which is explained in the next paragraph.

In order to obtain the offset sizes
  ($\Delta X_{j}^{\rm (QUAD)}$, $\Delta Y_{j}^{\rm (QUAD)}$,
   $\Delta X_{j}^{\rm (BPM)}$ and $\Delta Y_{j}^{\rm (BPM)}$),
  we perform 
    a global fitting for different orbits within a time period
    in which magnets are not expected to have moved.
In case of KEKB,
  we make various measurements in machine studies
  using six different orbits in each of the horizontal and vertical directions,
  which are produced by making a small kick
    at one of the six steering magnets.
We fit those six orbits simultaneously with
  not only the usual floating parameters
    ($X_i^{(in)},Y_i^{(in)},X_i^{(out)},Y_i^{(out)}$ ($i=1,2,3,4,5,6$)),
  but also the offsets in formula~(\ref{E:OrbitCalc:chi2:offsets}).
We furthermore introduce an additional floating parameter:
  the correction factor of the field strength of the bending magnet (BEND), $a$,
  which is defined as $\theta_{corr} = (1+a) \times \theta_{org}$,
  where $\theta_{org}$ and $\theta_{corr}$ are the kick angles
  before and after the correction, respectively.
The BPM and QUAD offsets as well as the correction factors of the BENDs 
  are common among the six orbits.
It is clear that the simple fitting procedure leads to failure
  because there are
    too many floating parameters (more than one hundred in our example).

Therefore,
  we first categorize the offset and correction-factor parameters to be floating
  into three sets:
  the BPM-related, QUAD-related, and BEND-related ones.
We choose zero or one parameter to be floating in each category, and
  the other parameters are fixed at zero.
Then, a global fitting is performed for the combination of the choice.
We try all of the possible combinations (about $10$ million in our example),
  and
  a large number of results are filtered, requiring the following criteria:
\begin{itemize}
  \item the minimization of $\chi^2$ converges
           with a positive-definite Hessian matrix,
  \item $|\Delta X_{j}^{\rm (QUAD)}| <10$\,mm
        and $|\Delta Y_{j}^{\rm (QUAD)}| <10$\,mm,
  \item $|\Delta X_{j}^{\rm (BPM)}| <10$\,mm
        and $|\Delta Y_{j}^{\rm (BPM)}| <10$\,mm,
  \item $| a |$ $<0.2$,
  \item the vertical and horizontal displacements of the fitted orbit
        at the IP with respect to the nominal IP are within $1$\,mm.
\end{itemize}
Then, the best (lowest) $\chi^2$ sample is selected as the final one.
Figure~\ref{F:OrbitCalc:fit_final} shows the fitted orbit
  after the offset corrections.
With the corrections, the $\chi^2$ is small, and the orbit passes close to
the BPM measurements and the nominal IP.
Choosing two offsets to be floating in each category,
  we obtained almost the same results.

This method is to find a solution
  with a minimum modification
  within the scope of the numerical approach.

\begin{figure}
  \centering
  \includegraphics[width=42mm,clip]{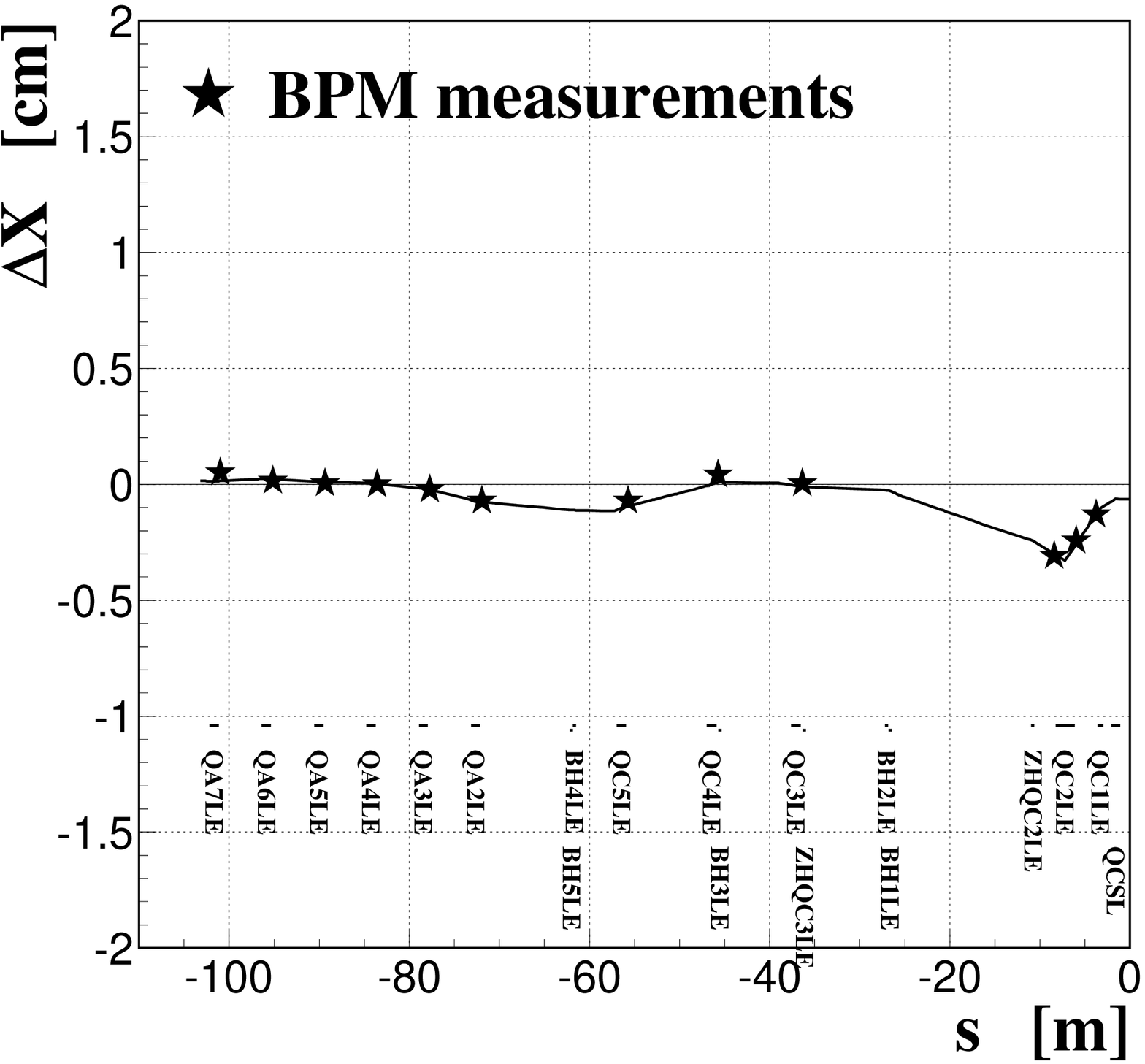}
  \includegraphics[width=42mm,clip]{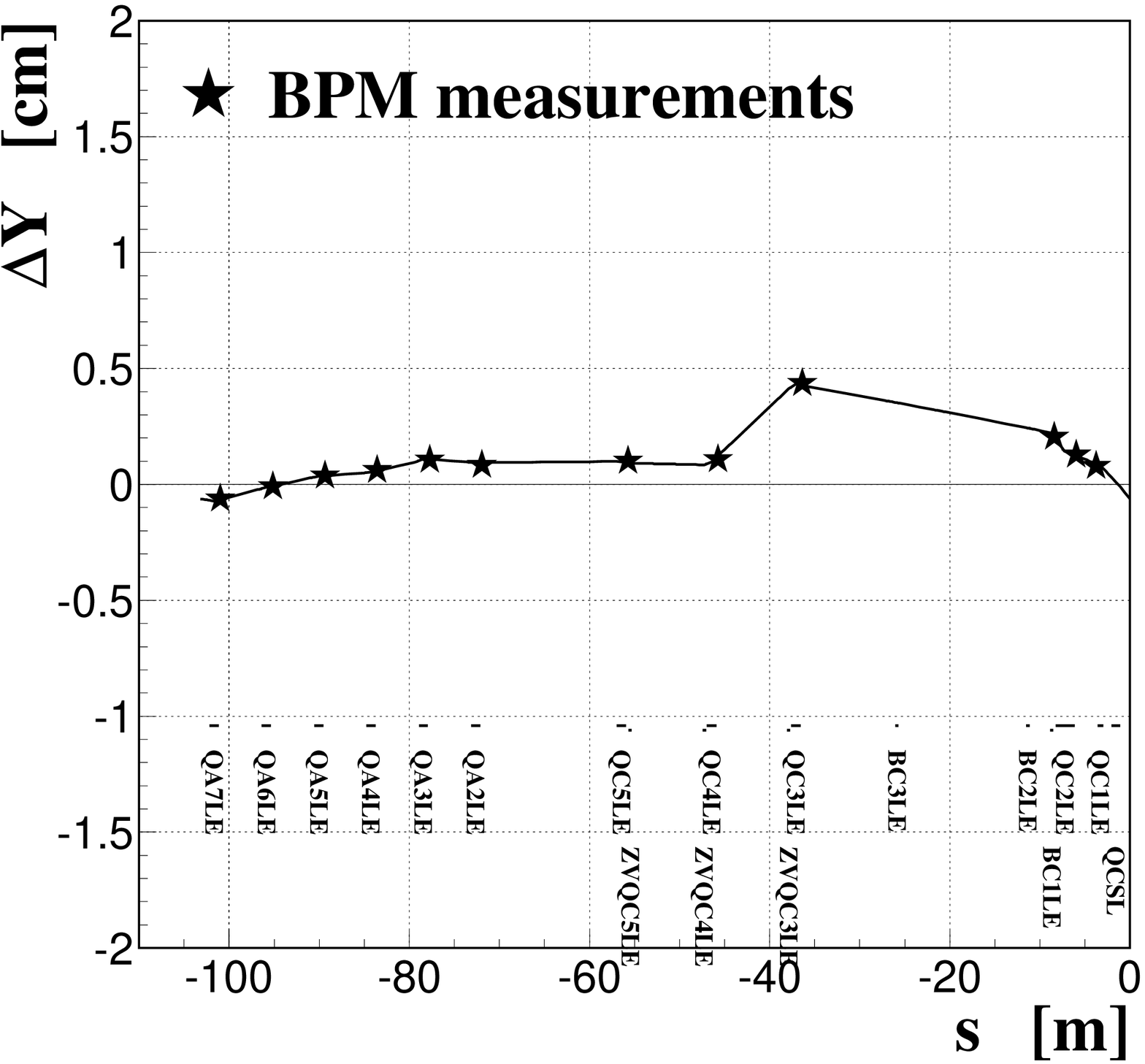}
  \caption{Example of the orbit-fitting results after the offset corrections.
           The left (right) figure shows the horizontal (vertical) fitted orbit
           together with the BPM measurements. 
           $\chi_X^2/N_{df(X)}=3.6$ and $\chi_Y^2/N_{df(Y)}=0.6$.
          } \label{F:OrbitCalc:fit_final}
\end{figure}


%% file: Wattage_calc/wattage.tex
\section{\label{R:Section:WattCalc}
  Wattage calculation}
\begin{figure}
  \centering
  \includegraphics[width=65mm,clip]{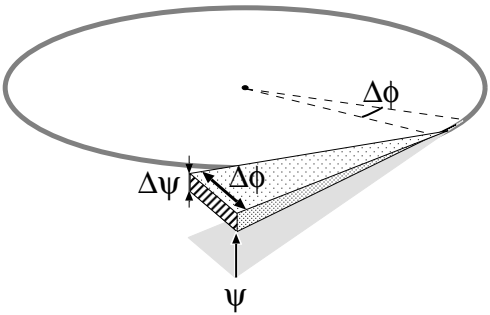}
  \caption{Integration variables: $\phi$ and $\psi$
           used in the wattage-calculation formula~(\ref{E:WattageCalc:Schwinger}).
           The circle indicates a circulating beam.
          } \label{F:WattageCalc:formula}
\end{figure}

SR wattages are calculated, based on the fitted real orbit,
  using the following analytical formula \cite{C:Schwinger:1949ym}:
\begin{eqnarray}
\Delta^3 W &=& \frac{3\alpha}{4\pi^2} \frac{I}{e} \gamma^2  \, 
       {\Delta\omega} {\Delta\phi} {\Delta\psi}
           \,\frac{1}{\omega}
           \left( \frac{\omega}{\omega_c} \right)^2
           \left( 1+\xi^2 \right)^2  \nonumber \\
       & & ~~~ \times
           \left\{ K_{2/3}(\eta)^2
                 + \frac{\xi^2}{1+\xi^2} K_{1/3}(\eta)^2
           \right\}
   \hbar \omega  ~{\rm ,}
   \label{E:WattageCalc:Schwinger}
\end{eqnarray}
  where
    $\alpha$ indicates the fine-structure constant,
    $I$ the beam current,
    $e$ the charge of the beam particle,
    $\gamma$ the Lorentz factor,
    $\omega$ the angular frequency,
    $\omega_c$ the critical angular frequency,
    $\phi$ the bending angle,
    $\psi$ the angle to describe the deviation from the bending plane,
    $\xi \equiv \gamma \psi$,
    $\eta \equiv (1/2)(\omega/\omega_c) \left(1+\xi^2\right)^{3/2}$,
    $\hbar$ Planck constant (reduced),
    and $K_{2/3}$ and $K_{1/3}$ are modified Bessel functions.
Figure~\ref{F:WattageCalc:formula}
  explains the two variables:
    $\phi$ and $\psi$.
%
We regard $\Delta \phi$ as being an infinitesimal quantity,
  and the integral form for formula (\ref{E:WattageCalc:Schwinger}) is 
  approximated by the following three-dimensional integration:
\begin{eqnarray}
W &=& \frac{3\alpha}{4\pi^2} \frac{I}{e} \gamma^2
    \int\hspace{-1.2ex} \int\hspace{-1.2ex} \int  \hspace{-0.6ex}
       {d\omega} {ds} {d\psi}
           \,\frac{1}{\omega}
           \left( \frac{\omega}{\omega_c} \right)^2 
           \left( 1+\xi^2 \right)^2  \nonumber \\
    & & ~~~ \times
           \left\{ K_{2/3}(\eta)^2
                 + \frac{\xi^2}{1+\xi^2} K_{1/3}(\eta)^2
           \right\}
   \frac{\hbar\omega}{\rho(s)}
   ~{\rm ,}
   \label{E:WattageCalc:integ}
\end{eqnarray}
  where the $s$ axis is taken to be along the beam orbit, and
  $\rho(s)$ is the radius of curvature at $s$.
A wattage distribution in a plane can be obtained
  by taking the mapping of $(s,\psi) \rightarrow (X,Y)$,
  where $(X,Y)$ is a point of intersection of the plane and
  the line with the angle $\psi$ with respect to
    the tangent of the orbit at $s$.
Integration on $s$ and $\psi$ corresponds to
  surface integration of the Poynting vector in the plane.
We perform numerical integrations
  of formula~(\ref{E:WattageCalc:integ})
  using the Monte-Carlo (MC) integration program
    BASES \cite{C:BasesSpring:1995th}.

The implementation of the wattage calculation in our program
  was checked by comparing with the numerical calculation using SAD
  \cite{C:KEKB:SAD:speed}.
In the SAD calculation, the exact formula in electrodynamics
  \cite{C:KEKB:SAD:E} is used.


It is important to consider any beam-size effects in the SR simulation,
  especially for the SR from QUADs.
The effects are included
  by adding four additional integration variables $(X,X',Y,Y')$,
  which indicate a phase-space point of the orbit at the IP.
MC integration is performed
  together with the following Gaussian weight function
  to describe the beam profile:
\begin{eqnarray}
  \frac{1}{ (\sqrt{2\pi})^4
     \sigma_{X}\sigma_{X'}\sigma_{Y}\sigma_{Y'}}
  &\times& \exp\left\{ -
     \frac{\left(X-X_{fit}\right)^2}{2\sigma_{X}^2}
             \right\} \nonumber \\
  &\times& \exp\left\{ -
     \frac{\left(X'-X'_{fit}\right)^2}{2\sigma_{X'}^2}
             \right\} \nonumber \\
  &\times& \exp\left\{ -
     \frac{\left(Y-Y_{fit}\right)^2}{2\sigma_{Y}^2}
             \right\} \nonumber \\
  &\times& \exp\left\{ -
     \frac{\left(Y'-Y'_{fit}\right)^2}{2\sigma_{Y'}^2}
             \right\} ~{\rm ,}
\end{eqnarray}
  where
  $(X_{fit},X_{fit}',Y_{fit},Y_{fit}')$ indicates the phase-space point
    of the fitted orbit at the IP,
  $\sigma_X$ ($\sigma_Y$) is the horizontal (vertical) beam size, 
  and
  $\sigma_{X'}$ and $\sigma_{Y'}$ indicate the beam size in $X'$-$Y'$ space.
The beam sizes are calculated according to the following formulas:
\begin{eqnarray}
  \sigma_X    &=& \sqrt{\epsilon_X \beta_X^*} ~{\rm ,} \\
  \sigma_{X'} &=& \sqrt{\epsilon_X/\beta_X^*} ~{\rm ,} \\
  \sigma_Y    &=& \sqrt{\epsilon_Y \beta_Y^*} ~{\rm ,} \\
  \sigma_{Y'} &=& \sqrt{\epsilon_Y/\beta_Y^*} ~{\rm ,} 
\end{eqnarray}
  assuming $\alpha^* =0$,
  where
    $\alpha^*$ is one of the Courant-Snyder parameters at the IP
    in the notation: $\gamma x^2 + 2\alpha x x' + \beta {x'}^2 = \epsilon$.
In the above formulas, $\epsilon_X$ ($\epsilon_Y$) indicates
  the horizontal (vertical) emittance, and
  $\beta_X^*$ ($\beta_Y^*$) is the horizontal (vertical) beta function
  at the IP.
The orbits in the relevant magnets 
  are calculated using the inverse transfer matrices from the IP.
This implementation in our program has been checked by comparing
  the results with no beam size and
          with a beam size in the asymptotic behavior:
    $\epsilon_X \rightarrow 0$ and $\epsilon_Y \rightarrow 0$.

\begin{figure}
  \centering
  \includegraphics[width=60mm,clip]{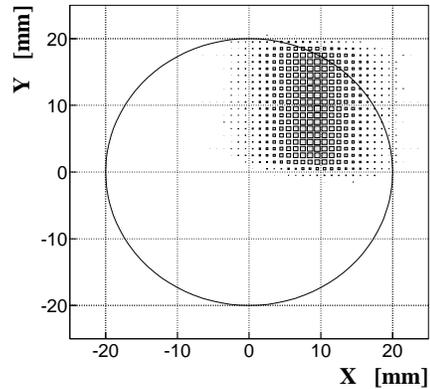}
  \caption{Example of the wattage distributions at the IP
           in the plane perpendicular to the beam axis.
           The $X$ axis is taken to be in the horizontal direction,
             pointing to the outside of the KEKB ring, and
           $Y$ in the vertical direction.
           The SR comes from the vertical BEND located about 25\,m upstream from the IP.
           The circle in the figure corresponds to the inner surface of the IP chamber.
          } \label{F:WattageCalc:BC3LE}
\end{figure}

In Fig.~\ref{F:WattageCalc:BC3LE},
  an example of the wattage distributions at the IP is shown
  in the plane perpendicular to the beam axis.
The SR in this example 
  was a dominant source in the SVD gain-drop accident.

%% file: Benchmark/btest.tex
\section{\label{R:Sec:BenchmarkTest}
         Bench-mark Test}
In order to demonstrate the validity of our method,
  we adopt the SVD gain-drop accident as a bench mark.
For radiation-dose calculations, 
  we made a detector simulation based on EGS4 \cite{Nelson:1985ec},
  which was improved for low-energy photons 
  \cite{Namito:1994rn,Namito:1997ya,Namito:1998,EGS4:KEKI00-3,EGS4:KEKI00-4}.
%
In this detector simulation, 
  the exact geometry and materials of the relevant SVD components
  have been installed according to engineering drawings.
SR photons, generated with
   the general-purpose event generator SPRING \cite{C:BasesSpring:1995th},
  are fed into the detector simulation, and
  energy deposits in the SVD are calculated.
Photons are simulated down to $1$\,keV, and
  electrons 
    to $20$-keV kinematic energy.
In order to transfer the radiation dose into the gain drop,
  we use the result on the irradiation test
  to investigate the radiation hardness of the SVD preamplifiers
    \cite{C:VA1:yokoyama}.

\begin{figure}
  \centering
  \includegraphics[width=64mm,clip]{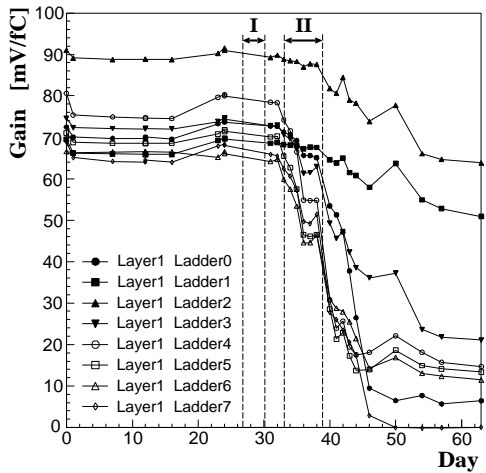}
  \caption{SVD gain as a function of time around the SVD gain-drop accident,
           together with the two periods used in the bench-mark test.
           The ladder numbers in this figure correspond to the position numbers
             in Tables~\ref{T:BenchMark:results:A}, \ref{T:BenchMark:results:B},
             and Fig.~\ref{F:BenchMark:gainresult}.
           The gain-drop plots are extracted from \cite{C:SVD2:TDR}.
          } \label{F:BenchMark:gaindrop}
\end{figure}

\begin{figure}
  \centering
  \includegraphics[width=42mm,clip]{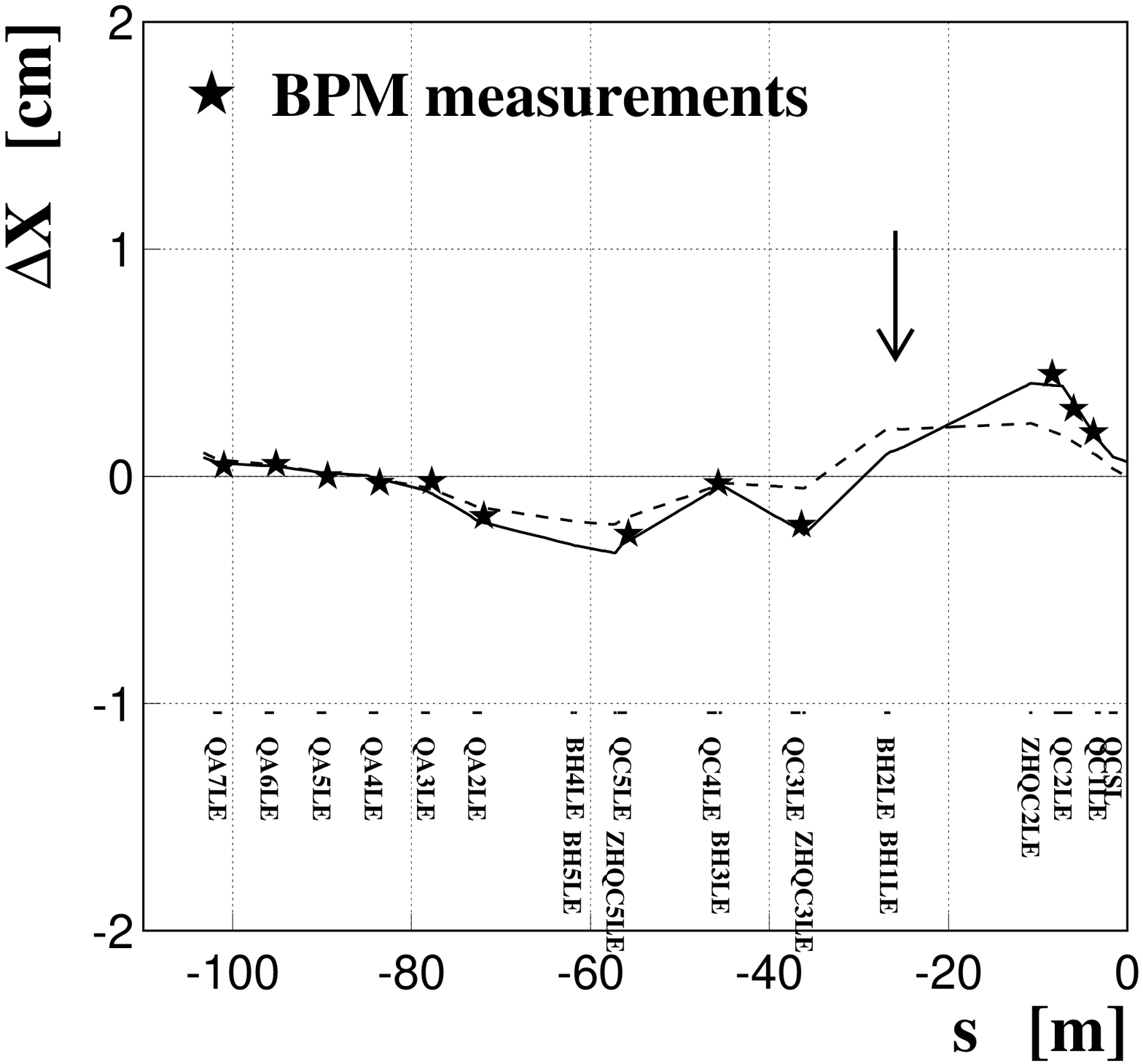}
  \includegraphics[width=42mm,clip]{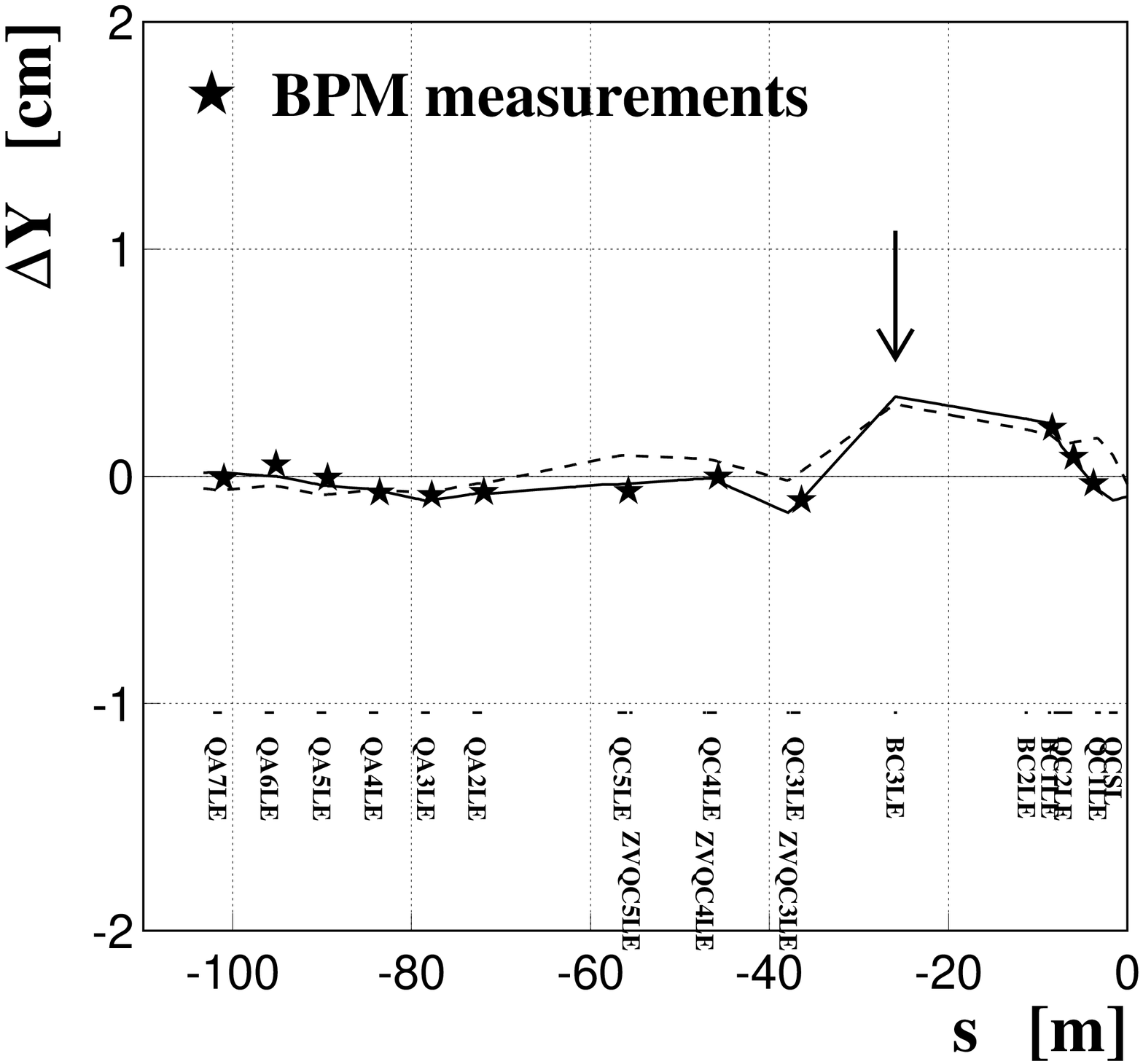}
  \caption{Fitted beam orbits around the SVD gain-drop accident.
           The dotted line indicates the orbit in period \Roman{one}
           and the solid line in period \Roman{two}.
           The vertical arrow points to the BEND which made a dominant contribution
           to the SVD gain drop in period \Roman{two}.
          } \label{F:BenchMark:orbits}
\end{figure}

\begin{figure}
  \centering
  \includegraphics[width=73mm,clip]{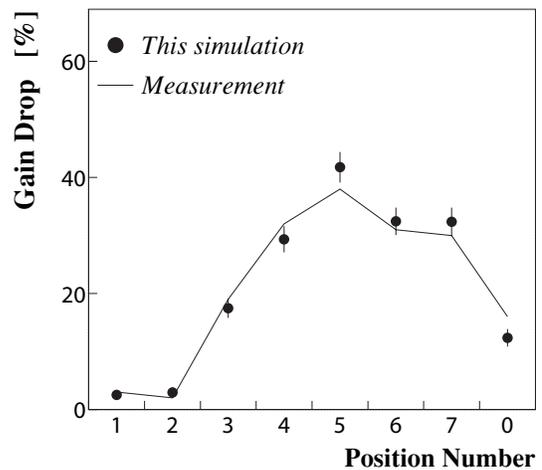}
  \caption{Comparison between the estimated and measured gain-drop values
           for period \Roman{two}.
          } \label{F:BenchMark:gainresult}
\end{figure}

We make two different sets of calculations for a more careful demonstration.
One of them is on the radiation dose
    before starting the gain drop (period \Roman{one}), and
  the other is after the start (period \Roman{two}),
  as shown in Fig.~\ref{F:BenchMark:gaindrop}.
During each of the two periods, the beam orbit was almost unchanged.
Table~\ref{T:BenchMark:results:A}} gives
  the results of a radiation-dose calculation for period \Roman{one}.
Obtaining information on the time of the period ($4641$ minutes),
  and on the average beam current ($4.6$\,mA),
  the gain drop is estimated to be around or less than $1$\,\%,
  which is consistent with the measurements ($\lesssim 1$\,\%).
After starting the gain drop,
  the estimated doses are much higher in period \Roman{two}
    than those in period \Roman{one},
  as shown in Table~\ref{T:BenchMark:results:B},
  where
  the time of the period is $8071$\,minutes,
  and the average beam current is $6.7$\,mA.
%
What changed significantly are
  the kick angle of the vertical BEND,
    which made a dominant contribution to the radiation dose 
    in period \Roman{two},
  and the orbit around the BEND, as shown in Fig.~\ref{F:BenchMark:orbits}.
The bending angle changed
  from $0.53$\,mrad with $1.7$-keV critical energy ($E_{crit}$)
  to $0.66$\,mrad with $E_{crit}=2.2$\,keV
  in the same direction.
%
Figure~\ref{F:BenchMark:gainresult} shows a comparison
  between the estimated and measured gain-drop values.
We have obtained good agreements not only on the absolute size,
  but also for the azimuthal distribution.

\begin{table*}
\begin{tabular}{lcccccccc}
  \hline
  \hline
    Position number  &  1  &  2  &  3  &  4  &  5  &  6  &  7  &  0 \\
  \hline
   Dose [Rad$/$min$/$A]
    & $2.1${\footnotesize $\pm 0.9$}
    & $1.2${\footnotesize $\pm 0.7$}
    & $23${\footnotesize $\pm 3$}
    & $41${\footnotesize $\pm 4$}
    & $33${\footnotesize $\pm 4$}
    & $33${\footnotesize $\pm 4$}
    & $25${\footnotesize $\pm 3$}
    & $3.5${\footnotesize $\pm 1.2$}
    \\
   Integrated dose [kRad]
    & $0.06$
    & $0.04$
    & $0.53$
    & $0.93$
    & $0.75$
    & $0.74$
    & $0.57$
    & $0.10$ \\
   Estimated gain drop\,[\%]
    & {$0.02$}
    & {$0.01$}
    & {$0.17$}
    & {$0.30$}
    & {$0.24$}
    & {$0.23$}
    & {$0.18$}
    & {$0.03$} \\
  \hline
  \hline
\end{tabular}
\caption{Results of a radiation-dose calculation
         based on our method
         for the period before starting the gain drop (period \Roman{one}).
  The position numbers correspond to the azimuthal coordinate with respect to the beam axis,
  and the numbers   2,     4,     6,          0
        indicate  the top, inside, bottom and outside
  of the KEKB ring, respectively.
  The errors from the MC statistics are shown beside the dose values.
}
\label{T:BenchMark:results:A}
\end{table*}

\begin{table*}
\begin{tabular}{lcccccccc}
  \hline
  \hline
    Position number  &  1  &  2  &  3  &  4  &  5  &  6  &  7  &  0 \\
  \hline
   Dose [$10^2$Rad$/$min$/$A]
   & $1.5${\footnotesize $\pm 0.4$}
   & $1.7${\footnotesize $\pm 0.5$}
   & $10${\footnotesize $\pm 1$}
   & $17${\footnotesize $\pm 1$}
   & $24${\footnotesize $\pm 2$}
   & $19${\footnotesize $\pm 1$}
   & $19${\footnotesize $\pm 1$}
   & $7.2${\footnotesize $\pm 0.9$}
\\
  Integrated dose [kRad]
   & $8.0$
   & $9.2$
   & $55$
   & $92$
   & $131$
   & $102$
   & $102$
   & $39$  \\
  Estimated gain drop [\%]
   & $2.6$
   & $2.9$
   & $17.6$
   & $29.4$
   & $41.9$
   & $32.6$
   & $32.6$
   & $12.5$  \\
  \hline
  \hline
\end{tabular}
\caption{Results of a radiation-dose calculation
       for the period after starting the gain drop (period \Roman{two}).
}
\label{T:BenchMark:results:B}
\end{table*}

%% file: conclusion.tex
\section{\label{R:Sec:conc}
  Conclusions and Future Prospects}
We have developed a new method to calculate SR wattages, or radiation doses,
  based on a real beam orbit
  with reasonable offset corrections.
Here,
  the orbit is obtained by fitting BPM measurements.
SR wattages are 
  calculated based on the fitted orbit with the MC integration
  of the analytical formula.
We have coded a dedicated computer program, where
  the correctness of the orbit and wattage calculations in our program
  has been confirmed by comparing with the results using SAD.
Finally,
  we have successfully reproduced the SVD gain-drop accident quantitatively,
  so that the practicability of our method has been established.

This method can be applied to the construction of an online alarm system
  for SR background,
  leading to prior notice for detector damage and
             possibly more flexibility in accelerator operation.
%
%
With this technique, a warning can be issued when the orbit is
    approaching dangerous areas, and furthermore the expected amount of
    SR background can be estimated before making certain changes
    in the orbit.

We have made
  a tentative version of
  the online program for KEKB operation.
It takes about twenty seconds per cycle (getting the magnet information and
  BPM measurements, fitting the orbit, and MC integration of SR wattages
  from all the magnets in the HER straight section near the IP)
  using the workstation with a 500MHz CPU.
If we use a recent faster computer,
  the turnaround time could be comparable
  to the time of BPM measurements (several seconds).

%% file: Appendix/transfer_matrices.tex
\section{\label{R:App:TransferMatrices}
  Transfer Matrices}
The exact forms of the transfer matrices used in the orbit calculation are
  shown in this appendix.
Here, $L$ indicates the effective length of the component.
$k$ is defined as $k=\sqrt{(B_0/b)/(B\rho)}$, where
  $B_0$ is the magnetic field strength at radius $b$,
  and $B\rho$ is the magnetic rigidity of the central reference trajectory.
$\theta_x$ ($\theta_y$) is the horizontal (vertical) kick angle
  in the BEND.
${\bf R}_{\rm rot}(\alpha)$ indicates
  the rotation matrix for the rotation angle, $\alpha$.
The elements with a blank mean zero.

{\scriptsize

\begin{eqnarray}
  {\rm Orbit}:~{\bf X}(s)
       \,=\,
      \left(
      \begin{array}{@{\,}l@{\,}}
        \hspace*{0.3ex} X(s) \\
        \hspace*{0.3ex} X'(s) \\
        \hspace*{0.3ex} Y(s) \\
        \hspace*{0.3ex} Y'(s) \\
        \hline 
        \hspace*{0.3ex} {\bf 1}
      \end{array}
         \right)
  \label{E:TransferMatrices:orbit}
\end{eqnarray}

\begin{eqnarray}
  {\rm Drift~space}:~
  {\bf R}_{\rm drift}(L) \,=\,
      \left(
      \begin{array}{@{\,}cccc|c@{\,}}
        1 & L &   &   &  \\
          & 1 &   &   &  \\
          &   & 1 & L &  \\
          &   &   & 1 &  \\
        \hline
          &   &   &   & {\bf 1} \\
      \end{array}
         \right)
  \label{E:TransferMatrices:drift}
\end{eqnarray}

\begin{eqnarray}
  {\rm QUAD(horizontal~focusing)}:~
  {\bf R}_{\rm quad} = \hspace{2.5cm} \nonumber \\
      \left(
      \begin{array}{@{\,}cccc|c@{\,}}
        \cos(kL) & k^{-1}\sin(kL) &   &   &  \\
       -k\sin(kL) & \cos(kL) &   &   &  \\
          &   & \cosh(kL) & k^{-1}\sinh(kL) &  \\
          &   & k\sinh(kL) & \cosh(kL) &  \\
       \hline
          &   &   &   & {\bf 1} \\
      \end{array}
         \right)
  \label{E:TransferMatrices:quad}
\end{eqnarray}

\begin{eqnarray}
  {\rm BEND}:
  {\bf R}_{\rm bend} =
      {\bf R}_{\rm drift}(\frac{L}{2}) \times
      \left(
      \begin{array}{@{\,}cccc|c@{\,}}
        1 &   &   &   &  \\
          & 1 &   &   & {\theta_x} \\
          &   & 1 &   &  \\
          &   &   & 1 & {\theta_y} \\
       \hline
          &   &   &   & {\bf 1}
      \end{array}
         \right)
      \times {\bf R}_{\rm drift}(\frac{L}{2})
  \label{E:TransferMatrices:bend}
\end{eqnarray}

\begin{eqnarray}
  {\rm Skew~QUAD}:
  {\bf R}_{\rm skewQ} =
      {\bf R}_{\rm rot}(-45^{\circ}) \times
      {\bf R}_{\rm quad} \times
      {\bf R}_{\rm rot}(+45^{\circ})
  \label{E:TransferMatrices:skew}
\end{eqnarray}

} 


%% file: srw.bbl
\begin{thebibliography}{17}
\bibitem[{C:K({\natexlab{b}})}]{C:KEKB:SAD:speed}
\bibinfo{note}{Typically, it takes ten hours to obtain a wattage distribution
  per a magnet using SAD, while using our program, shorter than one second.}
\bibitem[{C:K({\natexlab{c}})}]{C:KEKB:SAD:E}
\bibinfo{note}{SAD can calculate electric field produced by a point charge
  according to the Feynman-Heaviside formula which is derived from Maxwell's
  equations without any approximation.}
\end{thebibliography}
